\documentclass[%
 preprint,
 superscriptaddress,
 floatfix,
 amsmath,
 amssymb,
pra,
]{revtex4-1}

\usepackage{graphicx}
\usepackage{dcolumn}
\usepackage{bm}
\usepackage{longtable}
\begin{document}

\title{Electron affinities of At and its homologous elements Cl, Br, I}
\author{R. Si}
\affiliation{Department of Computer Science, University of British Columbia, Vancouver, Canada V6T 1Z4}
\author{C. Froese Fischer}\email{cff@cs.ubc.ca}
\affiliation{Department of Computer Science, University of British Columbia, Vancouver, Canada V6T 1Z4}

\begin{abstract}
The multiconfiguration Dirac-Hartree-Fock method is applied to study the electron affinities of At and its homologous elements Cl, Br and I.
Our method of calculation is validated through the comparison with the available experimental electron affinities of Cl, Br, I and other theoretical values.
The agreement between our predicted electron affinities and the available experimental values for Cl, Br and I is within 0.2\%, which is an improvement of more than a factor of 10 over previous theoretical studies.
Applying the same method to At, the electron affinity of At is predicted to be 2.3729(46)~eV.
\end{abstract}
\maketitle

\section{Introduction}
Astatine, element 85, is the rarest naturally occurring element on earth with an estimated total abundance of less than one gram~\cite{Asimov1953}.
One of its longest-lived isotopes, $^{211}$At, with a half-life of 7.2 hours, is of considerable interest as the most promising $\alpha$-particle emitting radionuclides for targeted radiotherapy~\cite{Bloomer1981,Zalutsky2008,Vaidyanathan2011,Gustafsson2012}.
The most important properties influencing its chemical behavior include the energy gained when an additional electron is attached to a neutral atom forming a negative ion, referred to as the electron affinity (EA).

Recently, the first electron affinity measurement of At was proposed but has not yet  been realized~\cite{Rothe2016}.
Although the electron affinity of At had been calculated through the use of various theoretical methods,
the reported theoretical values range from 2.110~eV to 3.183~eV~\cite{Sergentu2016,Borschevsky2015,Li2012,Laury2012,Chang2010,Zeng2010,Mitin2006,Roos2004,Peterson2003,Zollweg1969}.
Many methods are used -- some use numerically determined  orbitals obtained from a variational procedure, others an analytic basis,  some are based on the Dirac-Coulomb Hamiltonian and others are variants of the  Douglas-Kroll Hamiltonian~\cite{Wolf2002}, some add other relativistic or QED corrections.
Many of them were systematically calculated from Cl, Br and I to At, using the first 3 ions for which the experimental electron affinities are available to validate their approach.
However, their calculated electron affinities of Cl, Br and I all differ from the experimental values by over 2\%.
Here we also start from Cl but try to predict a more accurate EA value for At using a fully relativistic numerical approach.

\section{\label{sec:calculation}Calculation}
In the present work, the variational multiconfiguration Dirac-Hartree-Fock (MCDHF) method~\cite{Fischer2016} implemented in the GRASP2K package~\cite{Jonsson2013} is adopted.
The MCDHF method starts from a Dirac-Coulomb Hamiltonian $H_{DC}$
\begin{equation}
H_{DC}= \sum_{i=1}^N(c~\bm{\alpha_i}\cdot\bm{p_i}+(\beta_i-1)c^2+V_i)+\sum_{i>i}^N\frac{1}{r_{ij}},
\end{equation}
where $V_i$ is the monopole part of the electron-nucleus interaction for a finite nucleus, $r_{ij}$ is the distance between electrons $i$ and $j$, and $\alpha$ and $\beta$ are the Dirac matrices.  The nuclear density of a finite nucleus is a assumed to have a Fermi distribution function dependent on the mass of the isotope.
The electron correlation effect is included by expanding our Atomic State Function (ASF) $\Psi\left(\Gamma PJ\right)$ in a linear combination of Configuration State Functions (CSFs), $\Phi\left(\gamma_i PJ\right)$, namely
\begin{equation}
\Psi\left(\gamma PJ\right)=\sum_{i=1}^Mc_i\Phi\left(\gamma_i PJ\right),
\end{equation}
where $\gamma_i$ represents all other quantum numbers needed to uniquely define the CSF.
The CSFs are four-component spin-angular coupled, antisymmetric products of Dirac orbitals
of the form
\begin{equation}
\phi({\bf r})=\frac{1}{r}\left(\begin{array}{c}P_{n\kappa}(r)\chi_{\kappa m                            }(\theta,\phi)\\
iQ_{n\kappa}(r)\chi_{-\kappa m}(\theta,\phi)\end{array}\right).
\end{equation}
The radial parts of the one-electron orbitals and the expansion coefficients $c_i$ of the CSFs are obtained by the relativistic self-consistent field (RSCF) procedure.  Differential equations for the large and small components of the radial functions are determined from the variational principle for a stationary energy, stationary for all allowed perturbations.  The latter include perturbations to negative energy states that satisfy bound-state boundary conditions. The equations are solved numerically.

In the present paper,
the CSF expansions are obtained using the restricted active set (RAS) method~\cite{Olsen1988,Sturesson2007}, by allowing single and double (SD) substitutions from the reference configurations to an active orbital set.
The configurations with wave function compositions above 0.2\% are defined as our multireference (MR) set thereby including selected  triple (T) and quadrupole (Q) excitations relative to the initial CSFs.
In this study, electrons are divided into valence electrons and core electrons where only some of the latter are active in the SD process.
If SD excitations from only the valence electrons are allowed, we can include the valence-valence (VV) correlation effect; if we allow for excitations from one valence electron and one core electron, the core-valence (CV) correlation effect is taken into account; if double excitations from core electrons are allowed, we can include the core-core (CC) correlation effect.
To monitor the convergence of the calculated energies, the active sets are increased in a systematic way by adding layers of new orbitals.

The RSCF procedure determines the orbital basis.
In this work, we separated CSFs into two spaces, the CSFs that are members of the MR set along with those that include only VV correlation define the zero-order space $P$, while all other  CSFs  (remaining  CV or CC correlation)  define the first-order space $Q$.
With this separation,  a first-order  Hamiltonian interaction matrix ($H^{ZF}$) can be defined in terms of  four submatrices
\begin{equation}
H^{ZF}=\left(\begin{array}{cc}H^{(PP)} & H^{(PQ)}\\
H^{(QP)} & H^{(QQ)}\end{array}\right),
\end{equation}where all interactions within the zero-order space and between the zero- and first-order space submatrices are included, but only the diagonal energies are included in  $H^{(QQ)}$.  With this Hamiltonian,  the orbital basis is defined by interactions with significant components of the wave function or valence correlation basis and greatly reduces the time required for building a correlation basis.

Each RCSF calculation is followed by a Relativistic Configuration Interaction (RCI) calculation~\cite{McKenzie1980}, where the Dirac orbitals are kept fixed, and only the expansion coefficients of the CSFs were determined by finding selected eigenvalues and eigenvectors of the complete interaction matrix.
In this procedure, the transverse photon interaction  and  leading quantum electrodynamic (QED) effects (vacuum polarization and self-energy) were included, where the transverse photon interaction is included in the low-frequency limit
\begin{equation}
B_{ij}=-\sum_{i<j}^N\frac{1}{2r_{ij}}\left[\left(\bm \alpha_{i}\cdot \bm \alpha_{j}\right)
+\frac{\left(\bm\alpha_i\cdot\bm r_{ij}\right)\left(\bm\alpha_j\cdot\bm r_{ij}\right)}{r_{ij}^2}\right],
\end{equation}
the self-energy correction is obtained based on a screened hydrogenic approximation~\cite{Klarsfeld1973,Mohr1983}, the vacuum polarization correction is evaluated using the Uehling model potentials together with some higher order corrections~\cite{Fullerton1976}.
All calculations were performed with the GRASP2K code~\cite{Jonsson2013}.
The computational details are summarized in table~\ref{tab_MR} for each element and described in the following section.
\begin{table}
\setlength{\tabcolsep}{4pt}
\footnotesize
\centering
\caption{The summary of our computational details.\label{tab_MR}  Included is the mass number, type of correlation, the number of CSFs ($N_{CSF}$) in the expansion for $J=3/2$  (atoms) or $J=0$ (anion), and the configurations  in the MR set, all in non-relativistic notation. The notation CV$_{nl}$ indicates that core valence correlation includes an $nl$ core orbital, whereas CC$_{nl}$ indicates a double excitation from the $nl$ core subshell .}
\begin{tabular}{ l  r  l  r  l cccc}
\hline
    & Mass & Correlation  &    $N_{CSF}$ & MR set \\
\hline
Cl     & 35 & VV+CV$_{2p}$  & 171 957 & $3s^23p^5$, $3s^23p^33d^2$, $3s3p^53d$  \\
Cl$^-$ & 35 & VV+CV$_{2p}$ & 175 965 & $3s^23p^6$, $3s^23p^43d^2$, $3s3p^53d4f$, \\

       &    &              &         & $3s^23p^44p^2$, $3s3p^54s4p$, $3s3p^53d4p$  \\
Br     & 80 & VV+CV$_{3p,3d}$ & 292 360 & $3d^{10}4s^24p^5$, $3d^{10}4s^24p^34d^2$, \\
       &    &              &         &  $3d^{10}4s4p^54d$, $3d^{10}4s4p^54f$  \\
Br$^-$ & 80 & VV+CV$_{3p,3d}$& 164 975 & $3d^{10}4s^24p^6$, $3d^{10}4s^24p^44d^2$, \\
       &    &              &         &   $3d^{10}4s4p^54d4f$, $3d^{10}4s^24p^45p^2$  \\
I      & 127 & VV+CV$_{4p,4d}$+CC$_{4d}$ & 876 413 & $4d^{10}5s^25p^5$, $4d^{10}5s^25p^35d^2$, $4d^{10}5s5p^55d$, \\
       &     &    &         & $4d^84f^25s^25p^5$, $4d^94f5s^25p^45d$  \\
I$^-$  & 127 & VV+CV$_{4p,4d}$+CC$_{4d}$ & 440 018 & $4d^{10}5s^25p^6$, $4d^{10}5s^25p^45d^2$, $4d^{10}5s5p^55d5f$, \\
       &     &    &         & $4d^84f^25s^25p^6$, $4d^94f5s^25p^55d$, $4d^94f5s5p^66p$  \\
At     & 211 & VV+CV$_{5p,5d}$+CC$_{5d}$ & 930 502 & $5d^{10}6s^26p^5$, $5d^{10}6s^26p^36d^2$, $5d^{10}6s6p^56d$, \\
       &     &    &         & $5d^85f^26s^26p^5$, $5d^95f6s^26p^46d$  \\
At$^-$ & 211 & VV+CV$_{5p,5d}$+CC$_{5d}$ & 445 030 & $5d^{10}6s^26p^6$, $5d^{10}6s^26p^46d^2$, $5d^{10}6s6p^56d6f$, \\
       &     &    &         & $5d^85f^26s^26p^6$, $5d^95f6s^26p^56d$, $5d^95f6s6p^67p$  \\
\hline
\end{tabular}
\end{table}
\section{Results}
\subsection{Cl and Cl$^-$}
To find the strongly interacting CSFs for the ground state of Cl, we start from a tentative calculation, where the CSF list is generated by allowing the $3s$ and $3p$ electrons in the [Ne]$3s^23p^5$ configuration to be SD excited to $n\leq7, l\leq5$.
We notice that the CSFs with wave function compositions above 0.2\% are all from $3s^23p^5$, $3s^23p^33d^2$ and $3s3p^53d$ configurations, and the CSFs with wave function compositions above 0.1\% all arise from  configurations with $n\leq 4$ and $l\leq 3$.
Thus for Cl, $3s^23p^5$, $3s^23p^33d^2$ and $3s3p^53d$ configurations are chosen as our MR set.
Following the above computational strategy for Cl, we also did a similar tentative calculation for Cl$^-$.
It shows that Cl$^-$ is a much more complex system than Cl, the wave function compositions of some CSFs arising from $3s^23p^6$, $3s^23p^43d^2$, $3s3p^53d4f$, $3s^23p^44p^2$, $3s3p^54s4p$ and $3s3p^53d4p$ configurations are all above 0.2\%.
Thus these 6 configurations are included in the MR set for Cl$^-$.
In Cl and Cl$^-$, $2p$ is considered as an active core electron, $1s$, $2s$ are inactive core electrons, the other electrons in the MR configurations are considered as valence electrons.
VV correlation and CV correlation associated with the $2p$ electrons are included in our calculation.

For subsequent Cl calculations, the excitations of electrons in the MR configurations are generally restricted to $n\leq4, l\leq3$ except that valence electrons in the main configuration $3s^23p^5$ are allowed to be SD excited to $n\leq9, l\leq6$; one $2p$ electron and one valence electron are allowed to be SD excited to $n\leq9, l\leq5$.
The number of CSFs in the final $J=1/2$ and $J=3/2$ state expansions are 84 442 and 171 957, respectively. Only the largest expansion size is included in Table~\ref{tab_MR}.
The strategy of constructing the CSF list for Cl$^-$ is the same as for Cl, except for
the electrons in the MR set and the main configuration $3s^23p^6$ are allowed to be SD excited to $n\leq5, l\leq3$ and $n\leq10, l\leq5$, respectively.
The final expansion includes 175 965 CSFs.

The calculated fine-structure splitting of $3s^23p^5\ ^2P_{1/2,3/2}$ and the electron affinity of Cl are listed in table~\ref{tab_Cl}.
Our calculated fine-structure splitting agrees with the experimental value~\cite{NIST_ASD} by 0.3\%.
Two sets of EAs are provided, the first one, $E_n$(Cl)$-$$E_n$(Cl$^-$) (the $\Delta n=0$ EA), where $E_n$ labels the energy obtained when the ASF characterized by the maximum principal quantum number $n$ is used, the second one, motivated by the fact that more orbitals are needed to represent the negative ion, $E_{n-1}$(Cl)$-$$E_n$(Cl$^-$) (the $\Delta n=1$ EA).
Table~\ref{tab_Cl} shows that the former is increasing with $n$ whereas the latter is decreasing.
The $\Delta n=0$ and $\Delta n=1$ EAs can each be extrapolated approximately. Using the non-linear exponential decay function to extrapolate the last four $\Delta n=0$ and $\Delta n=1$ EA values to $n=\infty$, we get values of 0.13304(7)~a.u. and 0.13253(22)~a.u., respectively, where the digits in parentheses stand for the extrapolation error.
Taking an average of the two extrapolation values and adding the extrapolation errors together, we get an averaged extrapolation EA$=$0.13279(29)~a.u., this number differs from the experimental value 0.1327651(10)~a.u.~\cite{Berzinsh1995} by only 0.02\%.

\begin{table}
\setlength{\tabcolsep}{4pt}
\footnotesize
\centering
\caption{Fine-structure splitting (FS, cm$^{-1}$) of $3s^23p^5\ ^2P_{1/2,3/2}$ and electron affinity (EA, a.u.) of Cl from  Zero-First calculations.\label{tab_Cl}}
\begin{tabular}{cccccccc}
\hline
    & FS & E(Cl)   &   E(Cl$^-$) & EA ($\Delta n=0$) & EA ($\Delta n=1$) \\
\hline
$n$=4 & 854.03 & -460.958044 & -461.075931 & 0.117887 & 0.212341  \\
$n$=5 & 869.91 & -460.998711 & -461.126337 & 0.127626 & 0.168293  \\
$n$=6 & 878.37 & -461.013915 & -461.144706 & 0.130790 & 0.145994  \\
$n$=7 & 881.95 & -461.020169 & -461.152068 & 0.131899 & 0.138152  \\
$n$=8 & 883.74 & -461.022774 & -461.155125 & 0.132351 & 0.134956  \\
$n$=9 & 884.57 & -461.023918 & -461.156472 & 0.132554 & 0.133698  \\
$n$=10&        &             & -461.157207 &          & 0.133289  \\
Extrapolation& &           &             & \multicolumn{2}{c}{0.13279(29)} \\
Expt &882.3515\cite{NIST_ASD}&   &   & \multicolumn{2}{c}{0.1327651(10)\cite{Berzinsh1995}}    \\
\hline
\end{tabular}
\end{table}

\subsection{Br and Br$^-$}
The computational procedures for Br and Br$^-$ are similar as those for Cl and Cl$^{-}$.
The MR set for Br includes [Ar]$3d^{10}4s^24p^5$, $3d^{10}4s^24p^34d^2$, $3d^{10}4s4p^54d$ and $3d^{10}4s4p^54f$, while the MR set for Br$^-$ includes $3d^{10}4s^24p^6$, $3d^{10}4s^24p^44d^2$, $3d^{10}4s4p^54d4f$ and $3d^{10}4s^24p^45p^2$.
$3p$ and $3d$ are considered as active core electrons, $3s$ and $n=1,2$ electrons are inactive core electrons, the other electrons in the MR configurations are considered as valence electrons.
VV correlation and CV correlations associated with $3d$ and $3p$ electrons are taken into account.

For Br, valence electrons in the main configuration $3d^{10}4s^24p^5$ are allowed to be SD excited to $n\leq10, l\leq6$; one $3p$ or $3d$ electron and one valence electron in the main configuration are allowed to be SD excited to $n\leq10, l\leq5$; the excitations of electrons in the remaining MR set members are restricted to $n\leq4, l\leq3$.
For Br$^-$, electrons in the MR set, including the main configuration $3d^{10}4s^24p^6$,  can be excited to $n\leq5, l\leq3$ and $n\leq11, l\leq6$, respectively.
The number of CSFs in the final $J=1/2$ and $J=3/2$ state expansions for Br are 156 856 and 292 360, respectively. The final expansion list for Br$^-$ includes 164 975 CSFs.

The calculated fine-structure splitting of $3d^{10}4s^24p^5\ ^2P_{1/2,3/2}$ and electron affinity of Br are listed in Table~\ref{tab_Br}.
The fine-structure splitting differs from the experimental value by 0.16\%.
The extrapolation of the last four $\Delta n=0$ and $\Delta n=1$ EA values yield 0.12405(16)~a.u. and 0.12350(16)~a.u., respectively.
The averaged extrapolation value 0.12377(32)~a.u. differs from the experimental value 0.1236097(1)~a.u.~\cite{Blondel1989} by 0.13\%.

\begin{table}
\setlength{\tabcolsep}{4pt}
\footnotesize
\centering
\caption{Fine-structure splitting (FS, cm$^{-1}$) of $3d^{10}4s^24p^5\ ^2P_{1/2,3/2}$ and electron affinity (EA, a.u.) of Br.\label{tab_Br}}
\begin{tabular}{ccccccc}
\hline
    & FS & E(Br)   &   E(Br$^-$) & EA ($\Delta n=0$) & EA ($\Delta n=1$)  \\
\hline
$n$=5 & 3623.04 & -2603.100890 & -2603.213343 & 0.112453 & 0.221941  \\
$n$=6 & 3667.93 & -2603.142664 & -2603.261961 & 0.119297 & 0.161071  \\
$n$=7 & 3684.03 & -2603.157918 & -2603.279566 & 0.121649 & 0.136902  \\
$n$=8 & 3687.70 & -2603.164053 & -2603.286739 & 0.122687 & 0.128821  \\
$n$=9 & 3690.10 & -2603.166605 & -2603.289862 & 0.123257 & 0.125810  \\
$n$=10& 3690.45 & -2603.167726 & -2603.291216 & 0.123490 & 0.124612  \\
$n$=11&         &              & -2603.291917 &          & 0.124191  \\
Extrapolation& &             &              & \multicolumn{2}{c}{0.12377(32)}\\
Expt &3685.24\cite{NIST_ASD}&     &         & \multicolumn{2}{c}{0.1236097(1)\cite{Blondel1989}}          \\
\hline
\end{tabular}
\end{table}

\subsection{I and I$^-$}
For I, due to the fact that the $n=$4 shell is partially filled,
strong interactions are found from SD excitations involving
$4d\rightarrow4f$ and particularly $4d^2\rightarrow4f^2$ two-electron excitation.
The CSFs with wave function compositions above 0.2\% include not just
those generated by excitations from valence electrons of the main configuration
[Kr]$4d^{10}5s^25p^5$ such as $4d^{10}5s^25p^35d^2$ and $4d^{10}5s5p^55d$,
but also configurations like $4d^84f^25s^25p^5$ and $4d^94f5s^25p^45d$.
The above 5 configurations are included in the MR set of I.
For I$^-$, the MR set includes $4d^{10}5s^25p^6$, $4d^{10}5s^25p^45d^2$,
$4d^{10}5s5p^55d5f$, $4d^84f^25s^25p^6$, $4d^94f5s^25p^55d$ and $4d^94f5s5p^66p$.
Here $4p$ and $4d$ electrons are treated as active core electrons,
$4s$ and the $n=1,2,3$ electrons are inactive core electrons, the other electrons in the MR configurations are treated as valence electrons.
VV correlation, CV correlation associated with $4p$ and $4d$ electrons, and CC correlation associated with $4d$ electrons are taken into account.
Note that $4d$ and $4f$ electrons in reference configurations $4d^84f^25s^25p^5$, $4d^94f5s^25p^45d$ and $4d^84f^25s^25p^6$, $4d^94f5s^25p^55d$, $4d^94f5s5p^66p$ are kept inactive.

For I, valence electrons and $4d$ electrons in the main configuration $4d^{10}5s^25p^5$ are allowed to be SD excited to $n\leq11, l\leq6$ and $n\leq11, l\leq5$, respectively; one $4p$ electron and one valence electron in the main configuration are allowed to be SD excited to $n\leq11, l\leq5$; the excitations of electrons in the MR set are restricted to $n\leq5, l\leq3$.
Here except for the CSFs generated by only valence electron excitations, the CSFs that arise from the $4d^84f^25s^25p^5$ and $4d^94f5s^25p^45d$ configurations are also included in the zero order space of the RSCF process.
The final expansions for $J=1/2$ and $J=3/2$ include 455 801 and 876 413 CSFs, respectively.

For I$^-$, electrons in the MR set and the main configuration $4d^{10}5s^25p^6$ are allowed to be SD excited to $n\leq6, l\leq3$ and $n\leq12, l\leq6$ respectively.
In addition to the CSFs generated by only valence electron excitations, the CSFs that arise from $4d^84f^25s^25p^5$, $4d^94f5s^25p^55d$ and $4d^94f5s5p^66p$ configurations are also included in the zero order space of the RSCF process.
The final expansion includes 440 018 CSFs.

The calculated fine-structure splitting of $4d^{10}5s^25p^5\ ^2P_{1/2,3/2}$ and electron affinity of I are listed in Table~\ref{tab_I}.
The fine-structure splitting agrees with the experimental value~\cite{NIST_ASD} by 0.11\%.
Extrapolating the last four $\Delta n=0$ and $\Delta n=1$ EA values, we get 0.11269(3)~a.u. and 0.11248(6)~a.u., respectively.
The averaged extrapolation 0.11258(9)~a.u. differs from the latest experimental EA~\cite{Rothe2017} by 0.15\%.

\begin{table}
\footnotesize
\centering
\caption{Fine-structure splitting (FS, cm$^{-1}$) of $4d^{10}5s^25p^5\ ^2P_{1/2,3/2}$ and electron affinity (EA, a.u.) of I.\label{tab_I}}
\begin{tabular}{ccccccc}
\hline
    & FS & E(I)   &   E(I$^-$) & EA ($\Delta n=0$) & EA ($\Delta n=1$) \\
\hline
$n$=6 & 7603.03 & -7107.415598 & -7107.515199 & 0.099601 & 0.257587 \\
$n$=7 & 7611.62 & -7107.460497 & -7107.568555 & 0.108058 & 0.152957 \\
$n$=8 & 7617.17 & -7107.475305 & -7107.585965 & 0.110659 & 0.125468 \\
$n$=9 & 7613.14 & -7107.481447 & -7107.593118 & 0.111672 & 0.117813 \\
$n$=10& 7612.16 & -7107.484072 & -7107.596186 & 0.112114 & 0.114739 \\
$n$=11& 7611.21 & -7107.485327 & -7107.597645 & 0.112319 & 0.113573 \\
$n$=12&         &              & -7107.598432 &          & 0.113105 \\
Extrapolation& &            &               & \multicolumn{2}{c}{0.11258(9)} \\
Expt& 7602.970(5)\cite{NIST_ASD}&           &  & \multicolumn{2}{c}{0.1124181(14)\cite{Rothe2017}} \\
    &                           &           &  & \multicolumn{2}{c}{0.11241788(1)\cite{Pelaez2009}} \\
    &                           &           &  & \multicolumn{2}{c}{0.1124176(4)\cite{Hanstorp1992}} \\
\hline
\end{tabular}
\end{table}

\subsection{At and At$^-$}
Following the computational procedures of I and I$^{-}$, the MR set for At includes configurations [Xe]$(4f^{14})5d^{10}6s^26p^5$, $5d^{10}6s^26p^36d^2$, $5d^{10}6s6p^56d$, $5d^85f^26s^26p^5$ and $5d^95f6s^26p^46d$, while the MR set for At$^-$ includes $5d^{10}6s^26p^6$, $5d^{10}6s^26p^46d^2$, $5d^{10}6s6p^56d6f$, $5d^85f^26s^26p^6$, $5f6s^26p^56d$ and $5f6s6p^67p$.
Here $5p$ and $5d$ electrons are treated as active core electrons,
$5s$ and the $n=1,2,3,4$ electrons are inactive core electrons, the other electrons in the MR configurations are treated as valence electrons.
VV correlation, CV correlation associated with $5p$ and $5d$ electrons, and CC correlation associated with $5d$ electrons are taken into account.
Note that $5d$ and $5f$ electrons in reference configurations $5d^85f^26s^26p^5$, $5d^95f6s^26p^46d$ and $5d^85f^26s^26p^6$, $5f6s^26p^56d$, $5f6s6p^67p$ are kept inactive. The role of $5g$ orbitals was much smaller than that of $5f$ with an generalized occupation number of about 0.004 compare to 0.026 for $5f$.

For At, valence electrons and $5d$ electrons in the main configuration $5d^{10}6s^26p^5$ are allowed to be SD excited to $n\leq12, l\leq6$ and $n\leq12, l\leq5$, respectively; one $5p$ electron and one valence electron in the main configuration are allowed to be SD excited to $n\leq12, l\leq5$; the excitations of electrons in the MR set are restricted to $n\leq6, l\leq3$.
The strategy of constructing the CSFs list for At$^-$ are the same as for At, except for electrons in the MR set and the main configuration $4d^{10}5s^25p^6$ are allowed to be SD excited to $n\leq7, l\leq3$ and $n\leq13, l\leq6$ respectively.
The final expansion lists for $J=1/2$ and $J=3/2$ states of At include 484 165 and 930 502 CSFs, and for $J=0$ state of At$^-$ include 445 030 CSFs.

The calculated fine-structure splitting of $5d^{10}6s^26p^5\ ^2P_{1/2,3/2}$ and electron affinity of At are listed in Table~\ref{tab_At}.
The fine-structure splitting is 23099.49~cm$^{-1}$.
The extrapolation of the last four $\Delta n=0$ and $\Delta n=1$ EA values predict 0.08719(5)~a.u. and 0.08722(12)~a.u., respectively, and the averaged value is 0.08720(17)~a.u..

\begin{table}
\footnotesize
\centering
\caption{Fine-structure splitting (FS, cm$^{-1}$) of $5d^{10}6s^26p^5\ ^2P_{1/2,3/2}$ and electron affinity (EA, a.u.) of At.\label{tab_At}}
\begin{tabular}{cccccccc}
\hline
    & FS & E(At)   &   E(At$^-$) & EA ($\Delta n=0$) & EA ($\Delta n=1$)\\
\hline
$n$=7 & 23152.52 & -22863.08891 & -22863.16562 & 0.076707 & 0.212445 \\
$n$=8 & 23113.23 & -22863.12082 & -22863.20511 & 0.084294 & 0.116203 \\
$n$=9 & 23113.08 & -22863.13046 & -22863.21657 & 0.086114 & 0.095751 \\
$n$=10& 23101.26 & -22863.13396 & -22863.22060 & 0.086645 & 0.090145 \\
$n$=11& 23102.12 & -22863.13556 & -22863.22249 & 0.086927 & 0.088532 \\
$n$=12& 23099.49 & -22863.13648 & -22863.22353 & 0.087048 & 0.087969 \\
$n$=13&          &              & -22863.22420 &          & 0.087717 \\
Extrapolation& &            &               & \multicolumn{2}{c}{0.08720(17)}\\
\hline
\end{tabular}
\end{table}

\subsection{Comparison with experimental and other theoretical values}
\begin{figure}
  \centering
  \includegraphics[width=\columnwidth]{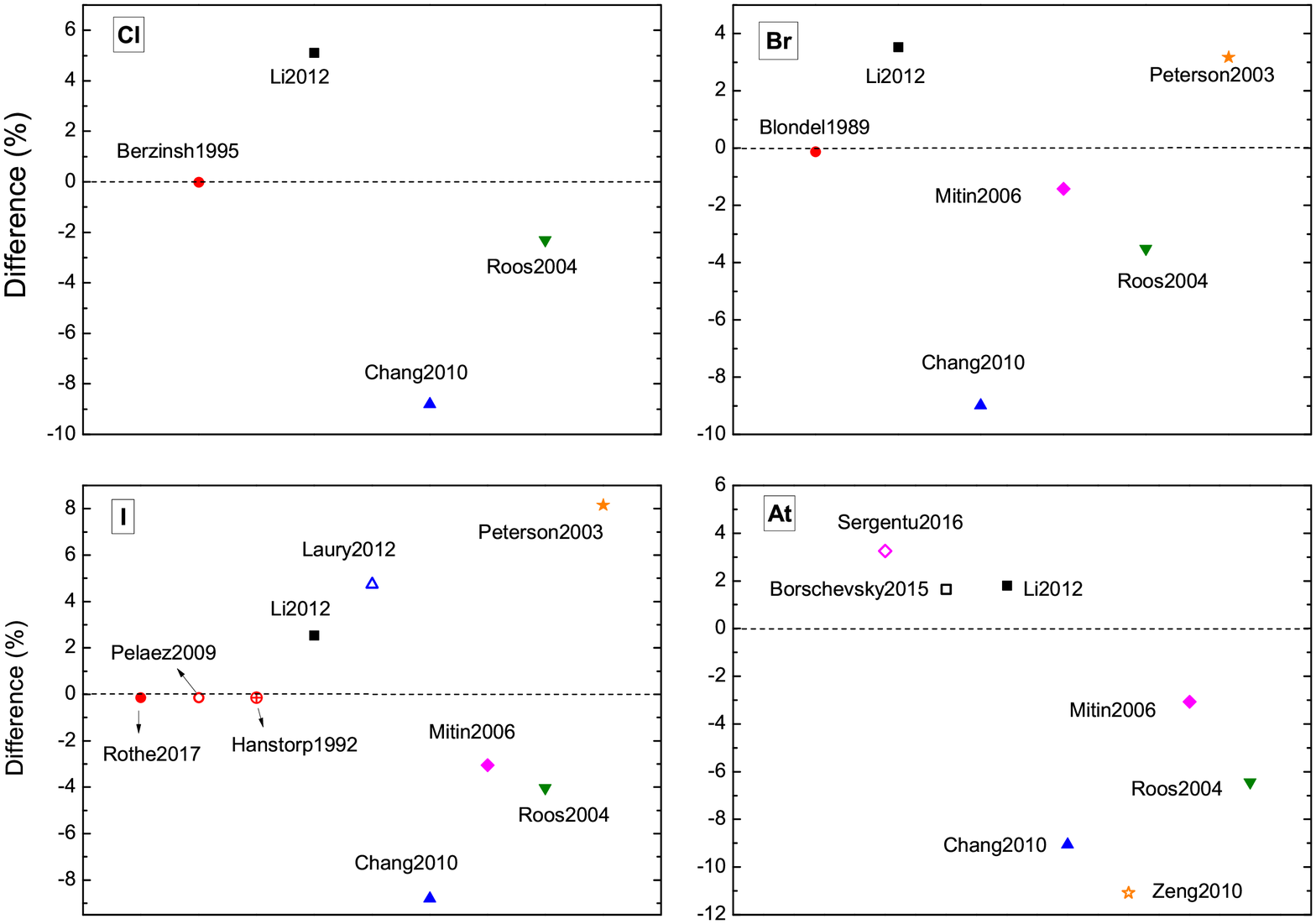}
  \caption{The electron affinities for Cl, Br, I and At from the present calculations are compared with the experimental and other theoretical values. The horizon lines stand for the present extrapolation values. See table~\ref{tab_EXP} for abbreviations of the references.}\label{fig_EXP}
\end{figure}

Our predicted EA values for Cl, Br, I and At (reported so far in a.u. or $E_h$ units) are compared with experimental and other theoretical values in Table~\ref{tab_EXP} and Figure~\ref{fig_EXP} in eV.

Experimental values, when available are accurate to 5-6 significant digits, accuracy that theory has not been able to match so far.
We can see from the Table~\ref{tab_EXP} and Figure~\ref{fig_EXP} that the present predicted EA values for Cl, Br and I all agree with the experimental values within 0.2\%. On the other hand, all the other theoretical values differ from the experimental values by over 2\%. At this point it should be pointed out that our uncertainties are uncertainties obtained from the extrapolation of computed energy differences.  All are small corrections.

In order to better understand the range of theory values we have grouped results by the Hamiltonian used in the calculation. The first category are results based on the Dirac-Coulomb Hamiltonian  plus  other corrections.  All three MCDHF results used essentially the same GRASP code~\cite{Jonsson2013,Parpia1996,Joensson2007}.  Li {\it et al.}~\cite{Li2012} included only valence correlation applying SD excitations to the main configuration of the atom and SDT to that of the anion. The effect of CV and CC was investigated and shown to reduce the EA.  Because of their SDT strategy for the anion, their computed EA is too large,  like our $\Delta n=1$ results.
Chang {\it et al.}~\cite{Chang2010} performed only SD excitations from the main configurations of the atom or anion which produced EA's that were too small, like our $\Delta n=0$ predictions.  In the case of At, they applied a semi-empirical extrapolation procedure for predicting EA from related elements. This extrapolation was not a small correction but one that changed the EA from 2.158~eV to 2.38~eV which is in good agreement with our present value of 2.3729(46)~eV.   All these three MCDHF calculations include the effect of a finite nucleus, the Breit and QED corrections.  The present results, which include some CV and CC results and  agree best with experimental values for homologous elements,  are the most reliable.

The results reported by Borschevsky {\it et al.}~\cite{Borschevsky2015} used a coupled-cluster CCSD(T) method where triple excitations are added as perturbative corrections with orbitals expressed in terms of an analytic uncontracted basis set.  Breit and QED were also added as a perturbative correction.  The final value 2.412~eV (or 0.08864 a.u.) is close to the present value but outside our uncertainty estimate.  The largest difference in the two strategies is the fact that they used a general SDT strategy for including higher-order effects whereas our procedure for the definition of the MR set is adaptive in that it selectively includes certain triple- or quadruple (TD) correlation effects, relative to the main configuration.  Also, their calculation was not a systematic calculation that  could be extrapolated.  Their value for the EA is close, in fact,  to the $\Delta n=1$ value of Table~\ref{tab_At} for $n=11$.

Most of the other theoretical results are based on variations of the Douglas-Kroll Hamiltonian~\cite{Wolf2002}, often a 2-component relativistic method, along with an analytic basis for orbitals.  No mention is made of the effect of a finite nucleus, Breit, or QED corrections.
The theoretical EA values from Mitin~{\it et al.}~\cite{Mitin2006} and Roos~{\it et al.}~\cite{Roos2004} are lower than our predictions by over 2\%.
The RECP+CCSD(T) EAs from Peterson~{\it et al.}~\cite{Peterson2003} for Br and I are overestimated by respectively 3\% and 8\%, and for At is higher than our prediction by 31\%.
The difference  is probably due to the omission  of CV/CC correlation and spin-orbit coupling effects.
Although the RECP-ccCA values from Laury~{\it et al.}~\cite{Laury2012} have included the above two effects, the EA for I is also overestimated by 5\%, and for At is larger than our prediction by 34\%.

\begin{table}
\footnotesize
\centering
\caption{The electron affinities (in eV) from the present calculations are compared with the experimental and other theoretical values.\label{tab_EXP} classified according to the Hamiltonian that was used.}
\begin{tabular}{llllllll}
\hline
Cl          &Br          &  I           &   At       & Method        & Ref.\\
\hline
3.612724(27)&            &              &            & Expt          & Berzinsh1995~\cite{Berzinsh1995}\\
            &3.3635880(20)&             &            & Expt          & Blondel1989~\cite{Blondel1989}\\
            &            & 3.059052(38) &            & Expt          & Rothe2017~\cite{Rothe2017}\\
            &            & 3.0590463(38)&            & Expt          & Pelaez2009~\cite{Pelaez2009}\\
            &            & 3.059038(10) &            & Expt          & Hanstorp1992~\cite{Hanstorp1992}\\
\multicolumn{6}{l}{1) Dirac-Coulomb Hamiltonian}\\
3.6133(79)  &3.3680(87)  & 3.0636(25)   & 2.3729(46)  & MCDHF+Extrap.  & Present \\
3.79807     &3.48677     & 3.14135      & 2.41591     & MCDHF          & Li2012~\cite{Li2012}\\
3.295       &3.065       & 2.794        & 2.158       & MCDHF          & Chang2010~\cite{Chang2010} \\
            &            &              & 2.38$\pm$0.02& MCDHF+Extrap. & Chang2010~\cite{Chang2010} \\
            &            &              & 2.412        & CCSD(T)       & Borschevsky2015~\cite{Borschevsky2015}\\
\multicolumn{6}{l}{2) Douglas-Kroll Hamiltonian  and others} \\
            &            &              & 2.45         & DFT           & Sergentu2016~\cite{Sergentu2016}\\
            &            & 3.2089       & 3.183        & RECP-ccCA     & Laury2012~\cite{Laury2012}\\

            &            &              & 2.110      & 4c-MBPT         & Zeng2010~\cite{Zeng2010}\\
            &3.32        & 2.97         & 2.30       & 2c-DK6+DFT      & Mitin2006~\cite{Mitin2006}\\
3.53        &3.25        & 2.94         & 2.22       & 2c-DK+CASPT2+SO & Roos2004~\cite{Roos2004}\\
            &3.475       & 3.313        & 3.115      & RECP+CCSD(T)    & Peterson2003~\cite{Peterson2003}\\
            &            &              & 2.8(2)     & Semi-empirical   & Zollweg1969~\cite{Zollweg1969}\\
\hline
\end{tabular}
\end{table}

It is interesting to note that, although the elements Cl, Br, I, and At are all group VIIA elements, the effect of correlation is  not the same for every element as shown in Table~\ref{tab_MR}.  As the $n$ of the $ns^2np^5$ configuration increases, the $(n-1)$ shell acquires more unfilled subshells.   For  I  ($n=5$),  CC became important because of the strong interaction from the $4d^2\rightarrow 4f^2$.  For At ($n=6$) , there are two unfilled subshells $5f, 5g$ but apparently only the $5d^2 \rightarrow  5f^2$ is strong, and so At is similar to I, but the role of $5g$  would increase for  Uus.

\section{Conclusion}
We report electron affinities of At and its homologous elements Cl, Br and I, by applying the MCDHF method.
Our calculated electron affinities for Cl, Br and I agree with the available experimental values within 0.2\%, which is a significant improvement over previous theoretical studies.
Using a similar computational approach, our prediction of the electron affinity of At is 2.3729(46) eV, in which the digits in the parentheses represent the extrapolation uncertainty.

\section*{Acknowledgement}
We acknowledge support from the NSERC Discovery Grant 2017-03851 of Canada and computer resources provided by Compute Canada.

\clearpage
\bibliographystyle{aip}
\bibliography{At}

\end{document}